\documentclass[12pt]{article}
\usepackage[T1]{fontenc}
\usepackage{amsmath}
\usepackage{amssymb} 
\usepackage{bbm} 
\usepackage[small]{caption2} 
\usepackage{fleqn} 
\usepackage{graphicx} 
\usepackage{mathrsfs}	
\usepackage[small,loose]{subfigure}  
\usepackage{cite} 
\usepackage{cancel} 
\usepackage{pifont} 
\usepackage{pstricks} 
\usepackage{longtable}
\usepackage{rotating}
\usepackage{nomencl} 
\makeglossary

\advance \headheight by 3.0truept       
\setcaptionwidth{0.75\textwidth}

\numberwithin{equation}{section}
\numberwithin{table}{section}

\allowdisplaybreaks

\begin{document}
\title{
\begin{flushleft}
{\normalsize CERN-PH-TH/2007-125}
\end{flushleft}
\vspace{2cm}
{\bf The Seesaw with Many Right--Handed Neutrinos }\\[0.8cm]}

\author{{\bf\normalsize 
John Ellis} and  {\bf\normalsize Oleg Lebedev }\\[1cm]
{\it\normalsize
TH Division, PH Department, CERN, CH-1211 Geneva 23, Switzerland }\\
}
\date{}
\maketitle \thispagestyle{empty} 

\abstract{
\noindent
There are no upper  limits on the possible number of massive, singlet (right--handed)
neutrinos that may participate in the seesaw mechanism, and some string constructions 
motivate seesaw models with up to ${\cal O}(100)$
right--handed neutrinos. In this case, the seesaw mass scale can be significantly {\it higher} than that
in the traditional scheme with just 3 right--handed neutrinos. 
We consider the possible phenomenological implications of such models, in particular, for 
lepton-flavour violation and electric dipole moments. 
Since the neutrino masses depend on the Majorana mass  scale linearly, while supersymmetric 
loop corrections depend on it logarithmically, the magnitude of lepton-flavour- and 
CP-violating transitions may {\it increase} with the multiplicity of the right--handed neutrinos 
and may be  enhanced by orders of magnitude. We also point out that, in the context of leptogensis, 
the bounds on the reheating temperature and the lightest
neutrino mass get  {\it relaxed} compared to those in the case of 3 right--handed neutrinos.}

\newpage

\section{Introduction}

The seesaw mechanism is arguably  the most attractive way to explain the smallness of 
neutrino masses \cite{Minkowski:1977sc}-\cite{Mohapatra:1979ia}.
In its conventional form, the  seesaw invokes 3 heavy singlet (right--handed) neutrinos $\nu_R$.
However, the number 3 is not sacred. On the one hand, the 2 non--zero light--neutrino mass
differences required by experiment could be explained with just 2 heavy right--handed neutrinos.
On the other hand, 3 non--zero masses could be explained with the participation of any
number $N \ge 3$ right--handed neutrinos. From the bottom--up perspective, there are no experimental constraints on the number of right--handed  neutrinos. For example, since they
are singlets of the Standard Model (SM) gauge group, their presence below the GUT
scale would not perturb the unification of the SM gauge couplings. It is certainly
possible, and even appears quite plausible, that such SM singlets do not follow the family pattern of 
the SM fermions.

Specific examples of scenarios with many right--handed  neutrinos are provided by
string models, and some recent string constructions motivate seesaw models with 
up to ${\cal O}(100)$ right--handed neutrinos \cite{Buchmuller:2007zd},\cite{Antoniadis:1987tv}. 
The reason is that string 
models contain abundant  SM singlets that often  have 
(non--renormalizable) couplings to the SM lepton doublets and  large Majorana masses, 
which are the  ingredients necessary for the seesaw mechanism.
This is just one of many scenarios that motivates a generalization of the conventional seesaw to
models with many right--handed neutrinos. 

In this Letter, we explore some of the phenomenological features  of such scenarios. Specifically, we comment that the seesaw mass scale may rise to {\it larger} values than in minimal schemes, that the magnitudes of lepton-flavour- and CP-violating effects may {\it increase}, and that leptogenesis
bounds on the reheating temperature and the lightest neutrino mass may be {\it relaxed}.

\section{Formulating the Seesaw with many Right--Handed Neutrinos}

The supersymmetric seesaw mechanism is
described by the superpotential (see e.g.\cite{Mohapatra:2006gs})
\begin{equation}\label{eq:seesawW}
 W~=~Y_e^{ij} \,\phi^d\,  e_i \ell_j \, + Y_\nu^{ij} \,\phi^u\,  N_i \ell_j \, + \frac{1}{2} M_{jk} N_j\, N_k \;,
\end{equation}
where $\phi^{u,d}$ and $\ell_i$ ($i=1,2,3$) are the Higgs and lepton doublets, 
$e_i$ are the charged SU(2)--singlet leptons 
and $N_j$
($1\le j\le n$) are some heavy Standard Model (SM) singlets with Majorana mass
terms  $M_{jk}$. The Yukawa couplings form an $n \times 3$ matrix, while 
the Majorana mass terms form an $n \times n$ matrix.
The resulting effective mass matrix for the left--handed neutrinos is given by 
\begin{equation}
 M_\mathrm{eff}~=~ - (v\sin\beta)^2 ~ Y_\nu^T\, M^{-1}\, Y_\nu  ~ \equiv ~ - {(v\sin\beta)^2 \over M_*}~ S \;,
\end{equation}
where  $v=174\,\mathrm{GeV}$, $\tan\beta$ is the ratio of the Higgs VEVs,  $M_*$ is the ``effective seesaw scale'' and
$S$ is a $3\times 3$ texture  whose largest entry is of order one.

As we now discuss, low-energy physics is quite sensitive to the number of right--handed neutrinos 
$n$ participating in the seesaw mechanism.
In particular, the effective seesaw mass scale $M_*$ may depend on $n$:
\begin{equation}
M_* ~\propto~ n^{-x} ~, 
\label{M}
\end{equation}
where $0 \leq x \leq 2$ is a function of the texture. We consider the following limiting cases
(in a particular basis defined, for example, by the Froggatt--Nielsen charges):

(1) all entries of $M$ and $Y_\nu$ contribute coherently to  $M_\mathrm{eff}$ with similar
magnitudes, so that $M_* ~\propto~ n^{-2}$; 

(2) the magnitudes of the different entries of  $M$, $Y_\nu$ are similar, but the 
complex phases are random\footnote{In this case, 
the dependence of $M_*$ on $n$  is obtained using the ``random walk'' result that
$$ \sum_{i=1}^N (-1)^{a_i}  \sim \sqrt{N}\;,$$
where $a_i$ is a random integer.}, so that $M_* ~\propto~ n^{-1}$;

(3) only a limited number of right--handed neutrinos provide significant contributions,
so that $M_* ~\propto~ n^{0} $. \\ \ \\
Typical realistic scenarios  correspond to intermediate 
situations. For instance, if $M$ is approximately diagonal with similar eigenvalues and 
all the $Y_\nu^{ij}$ are similar in magnitude but have arbitrary phases, $M_* ~\propto~ n^{-1/2}$.
As a specific example, we note that the stringy model of Ref.\cite{Buchmuller:2007zd}
corresponds to $x$ between 0 and 1.

In most cases, the masses of the left--handed neutrinos grow with $n$. Therefore,
in order to keep
them at the same values as in the conventional seesaw with 3 right--handed neutrinos,
one has either to {\it increase} the Majorana masses or to {\it decrease} the Yukawa couplings.
These possibilities differ in their phenomenological implications.

Consider, for simplicity, an MSSM scenario with universal soft supersymmetry-breaking scalar
masses at the GUT scale, as motivated by minimal supergravity (mSUGRA), complemented
by such a seesaw with $n$ right--handed neutrinos.
The renormalization-group (RG) running between the GUT scale and the scale at which the 
heavy neutrinos decouple that is due to the neutrino Yukawa couplings
induces additional  flavour--violating soft terms \cite{Borzumati:1986qx},
\begin{equation}
m_{\tilde l}^2 = \left( 
\begin{matrix}
m_L^2 & m_{LR}^{2\;\dagger} \\
m_{LR}^{2} & m_R^2 
\end{matrix}
\right)_{\rm mSUGRA} ~+~
\left(
\begin{matrix}
\delta m_L^2 & \delta m_{LR}^{2\; \dagger} \\
\delta m_{LR}^{2} & 0 
\end{matrix}
\right)\;,
\end{equation}
where $m_{\tilde l}^2$ is the slepton mass-squared matrix.
These corrections are given by \cite{Hisano:1998fj}-\cite{Petcov:2005jh}
\begin{eqnarray}
&& \delta m_L^2 \simeq -{1\over 8 \pi^2} (3m_0^2 + A_0^2) ~ Y_\nu^\dagger L Y_\nu \;, \nonumber \\
&& \delta m_{LR}^2 \simeq - {3 A_0 v \cos\beta \over 16 \pi^2} ~ Y_l Y_\nu^\dagger L Y_\nu \;.
\label{corr}
\end{eqnarray}
Here $L_{ij} \equiv \ln(M_{\rm GUT}/M_i)~ \delta_{ij}$ with $M_i$ being the heavy neutrino mass
eigenvalues. 

We first note that the dependence on the Majorana masses is only logarithmic. Thus,
increasing the scale of the Majorana masses does not affect these corrections significantly.
On the other hand,   $Y_\nu^\dagger  Y_\nu$ in general
grows  with the multiplicity of the heavy neutrinos, plausibly as  a power law. Therefore, 
ignoring the logarithmic piece, we have 
\begin{equation} 
\delta m_L^2~,~\delta m_{LR}^2 ~\propto ~ n^y \;,
\end{equation}
where the power $0 \leq y \leq 1$ depends on the 
texture. 
In the extreme case of $n$  ``coherent'' contributions,
$y=1$~\footnote{ Unlike in Eq.(\ref{M}), 
here we work in the basis where the Majorana mass matrix is diagonal. Thus, there are $n$ 
contributions in the sum which we assume to be similar in magnitude in this basis. 
Note that the basis change may,
in the case of many $\nu_R$'s, change the orders of magnitude of the couplings.}; 
for a case with random phases, $y=1/2$; finally,  $y=0$ when only a few heavy neutrinos contribute
significantly. Clearly, flavour--violating effects are expected to be enhanced, in general.
In particular, the branching ratio for radiative  $l_i \rightarrow l_j \gamma$
decays  \cite{Hisano:1998fj},\cite{Casas:2001sr},
\begin{equation}
\Gamma (l_i \rightarrow l_j \gamma) ~\propto~ \alpha^3 m_{l_i}^5 {  \vert (\delta m_L^2)_{ij}  \vert^2 
\over \tilde m^8  } \tan^2 \beta \;,
\end{equation}
where $\tilde m$ characterizes the typical sparticle masses in the loop
and $\alpha$ is the fine structure constant, is enhanced by $n^{2y}$ which can be a large factor
(up to 4 orders of magnitude)~\footnote{We also note that, since $Y_\nu^\dagger Y_\nu$ is larger 
than in the usual case,
the presence of many right--handed neutrinos improves the MSSM
gauge coupling unification at two loops \cite{Casas:2000pa}.}.

Alternatively, if instead of increasing the scale of the Majorana masses, one decreases the Yukawa
couplings, no enhancement of lepton--flavour--violating
(LFV) is expected, in general. This is because $M_{\rm eff} $ and $\delta m^2_{L,LR}$
both depend on $Y_\nu$ quadratically.

\section{Phenomenological Constraints}

The requirement of 
{\it perturbativity} restricts the magnitude of the loop corrections, which translates into
\begin{equation}
{(Y_\nu^\dagger  Y_\nu)_{ij} \over 4 \pi^2 } ~,~ {(Y_\nu^\dagger L   Y_\nu)_{ij} \over 4 \pi^2 }  
\leq {\cal O} (1) \;.
\end{equation}
This is a rather weak constraint and leaves open the possibility that some entries of 
$ Y_\nu^\dagger L  Y_\nu $ could be as large as 10.
We note that there are no (very) large logarithms in the loop corrections since the seesaw scale is 
quite high.

The main constraints are imposed by the {\it lepton--flavour--violating}
branching ratios \cite{Brooks:1999pu}-\cite{Aubert:2005ye}
\begin{eqnarray}
&& {\rm BR}(\mu \rightarrow e \gamma) < 1.2 \times 10^{-11} \;, \nonumber\\
&& {\rm BR}(\tau \rightarrow e \gamma) < 1.2 \times 10^{-7} \;, \nonumber\\
&& {\rm BR}(\tau \rightarrow \mu \gamma) < 6.8 ~(4.5) \times 10^{-8} \;. 
\end{eqnarray}
Supersymmetric flavour--violating effects can be expressed in terms of the ``mass insertions''
$\delta_{LR} \equiv \delta m^2_{LR} / m^2$ and $\delta_{LL} \equiv \delta m^2_{L} / m^2$ \cite{Hall:1985dx}, where
$m$ is the average slepton mass,
such that \cite{Gabbiani:1996hi}
\begin{equation}
 {\rm BR}(l_i \rightarrow l_j \gamma)= \Bigl\vert \xi_1 ~\delta_{LL}^{ij} + \xi_2 ~\delta_{LR}^{ij} \Bigr\vert^2 + (L \leftrightarrow R) \;.
\end{equation}
Here  $\xi_{1,2}$ are functions of SUSY masses whose explicit form is given in 
Ref.\cite{Gabbiani:1996hi}.
For electroweak--scale sparticle masses,
the resulting constraints on the mass insertions are \cite{Gabbiani:1996hi}:
\begin{eqnarray}
&& \bigl\vert \delta_{LL}^{12} \bigr\vert  < {\rm few}\times 10^{-3} ~~,~~ \bigl\vert \delta_{LR}^{12} \bigr\vert < 10^{-6} ~,\nonumber\\
&& \bigl\vert \delta_{LR}^{13} \bigr\vert  < 10^{-2} ~~,~~~~~~~~~\; \vert \delta_{LR}^{23} \vert < 10^{-2} ~. 
\end{eqnarray}
By means of Eq.(\ref{corr}), these bounds are translated at low $\tan\beta$ into
\begin{equation}
(Y_\nu^\dagger L Y_\nu)_{12} < 0.1 \;,
\end{equation}
with the other entries being essentially unconstrained.

The interpretation of this constraint depends strongly
on the magnitudes of the Yukawa couplings and the details of the texture. 
For smaller Yukawa couplings, the number of RH neutrinos can be very large.
However, for couplings of order unity, only a few RH neutrinos are allowed, 
unless there are  cancellations.
In general, even though $(Y_\nu^\dagger L Y_\nu)_{12}$ is rather small, other matrix 
elements can be significant and lead to observable effects.

\section{CP--Violating Phases and Electric Dipole Moments}

The Yukawa couplings and the Majorana mass terms are in general complex, leading 
in general to CP violation.
The number of physical CP--violating phases in the high--energy theory can be determined 
by parameter counting. Initially,
$Y_e$ has 9 complex phases, $Y_\nu$ has $3n$ phases, and $M$ has $n(n+1)/2$ phases. 
On the other hand, the ``flavour'' rotation symmetry 
\begin{equation}
U_l \times U_e \times U_N,
\end{equation}
acting as
\begin{eqnarray}
Y_e  & \rightarrow & U_e^\dagger ~ Y_e ~U_l ~,\nonumber\\
Y_\nu  & \rightarrow & U_N^\dagger ~ Y_\nu ~U_l ~,\nonumber\\
M  & \rightarrow & U_N^\dagger ~ M  ~U_N^* ,
\end{eqnarray}
allows one to eliminate some of these phases by field redefinitions. 
We note that $U_l$ and $U_e$ contain 6 phases each,
while $U_N$ has $n(n+1)/2$ phases~\footnote{In the pure Dirac case, one of these phases is 
irrelevant since 
an overall phase redefinition leaves all flavour objects invariant.}. 
Thus, this flavour rotation leaves $3 (n-1)$ physical phases, which
we can identify explicitly in a specific basis. Consider the basis where
\begin{eqnarray}
 Y^e &=& {\rm real~diagonal ~},\nonumber\\
 M &=& {\rm real~diagonal ~},\nonumber\\
 Y^\nu &=& {\rm arbitrary} \;.
\end{eqnarray}
This basis is defined only up to a diagonal phase transformation
\begin{equation}
U_l=U_e= {\rm diag}({\rm e}^{i\phi_1}, {\rm e}^{i\phi_2}, {\rm e}^{i\phi_3} ) \;,
\end{equation}
which acts on the neutrino Yukawa couplings as $Y_\nu^{ij} \rightarrow Y_\nu^{ij} \;  {\rm e}^{i \phi_j}$. 
The physical phases are invariant under this residual symmetry and are given by
\begin{equation}
{\rm Arg} \Bigl( Y_\nu^{ij} \; Y_\nu^{kj \; *} \Bigr) \;,
\label{invphases}
\end{equation}
(where no summation over $j$ should be understood).
Clearly, there are exactly $3(n-1)$ such phases, and each one may play the role of a ``Jarlskog invariant'' (see 
\cite{Dreiner:2007yz},\cite{Branco:1986gr}  for a discussion). 

Complex phases in the neutrino Yukawa couplings induce CP--violating phases in the soft terms
due to the RG running, which in turn contribute to lepton electric dipole moments (EDMs) at low 
energies \cite{Ellis:2001xt}-\cite{Farzan:2004qu}.
The relevant flavour objects  are the 3$\times$3 matrices
 $Y^\dagger_\nu Y_\nu$ and  $Y^\dagger_\nu L  Y_\nu$,
such that  \cite{Ellis:2001xt},\cite{Farzan:2004qu}
\begin{equation}
d_i \propto \Bigl[  Y^\dagger_\nu Y_\nu \; , \;  Y^\dagger_\nu L  Y_\nu    \Bigr]_{ii} \;,
\end{equation}
where $d_i$ is the EDM of the $i$-th charged lepton. 
This expression is proportional to Im $[Y_\nu^{jk} Y_\nu^{j'k \; *} $ $ \; Y_\nu^{j'i} Y_\nu^{ji\; *} L^{j'}] $ 
(no summation over $i$) which makes it clear that only the reparametrization--invariant phases 
(\ref{invphases}) are involved. Since the summation over $n$ RH neutrinos appears
twice,  the EDMs grow as
\begin{equation}
d_i \propto n^{2 y} \;.
\end{equation}
Thus, they  may be enhanced by several orders of magnitude in the multi--neutrino case.

At large $\tan\beta$, a different flavour structure appears \cite{Masina:2003wt},\cite{Farzan:2004qu},
namely:
\begin{equation}
d_i \propto  {\rm Im} \Bigl(  Y^\dagger_\nu f  Y_\nu ~ m_l^2 ~  Y^\dagger_\nu  g  Y_\nu    \Bigr)_{ii} \;,
\end{equation} 
where $f,g$ are diagonal matrices depending on the Majorana masses and  $m_l$ are the charged lepton masses.
The conclusion, however, remains the same, and the EDMs grow with the multiplicity of the states as 
$d_i \propto n^{2 y}$.

The order of magnitude of the induced EDMs can be estimated using  \cite{Farzan:2004qu}:
\begin{equation}
d_i \sim 10^{-29} \left(   { 200 ~{\rm GeV} \over  M_{\rm SUSY}}    \right)^2 
\left(   { m_{l_i} \over  m_e }    \right) (Y^\dagger_\nu Y_\nu)^2_{ii}  ~e~{\rm cm} \;,
\end{equation}
where we have taken  $\ln (M_i/M_j)= {\cal O}(1) $ and CP--violating phases that are of
order unity.
The current experimental limit on the electron EDM of  $10^{-27}$ $e$ cm is saturated for 
$(Y^\dagger_\nu Y_\nu)_{11} \sim 10$ if the superpartners of SM particles have masses
at the electroweak scale. For large $\tan\beta $, the above expression acquires an additional
factor $(\tan\beta /10)^3$ which leads to a further enhancement of the EDMs.

It is important to remember that these estimates are very sensitive to other CP--violating
phases.
For example, if the phase of the $\mu$--term is as small as $10^{-4}-10^{-5}$, 
it will dominate the SUSY
contributions to the EDMs (for a recent discussion, see \cite{Abel:2005er}).

\section{Numerical Example}

In this Section, we illustrate the multi--neutrino scenario with a numerical example.
We suppose that, in the basis where the charged lepton mass matrix is diagonal
and positive, 
the  mass matrix for the light neutrinos  
is given by
\begin{eqnarray}
&& M_{\rm eff}=\left(
\begin{array}{lll}
 \text{0.003} & \text{0.003} & \text{0.003} \\
 \text{0.003} & \text{0.028} & -\text{0.022} \\
 \text{0.003} & -\text{0.022} & \text{0.028}
\end{array}
\right) \nonumber
\end{eqnarray}
in eV units.  This mass matrix has eigenvalues $(0.05,0.01,0)$ eV and is diagonalized
by a tri-bimaximal  \cite{Harrison:2002er}  PMNS  transformation. 
In our convention, the (1,1) entry of $M_{\rm eff} $  
corresponds to the $\tau$--neutrino, (2,2) -- to the muon neutrino and so on.

We consider two seesaw realizations of this light-neutrino mass matrix: 
one with 10 right--handed neutrinos 
and another with 2  right--handed neutrinos. 
The corresponding heavy-neutrino mass matrices are generated randomly under the 
conditions that the Yukawa couplings be of order one  and 
that most of  the entries of $M^{-1}$ be similar in magnitude,
while reproducing
the correct $M_{\rm eff}$. For simplicity, we choose  real matrices. 

In the case of 10 right--handed neutrinos,
the inverse Majorana mass matrix $M^{-1}$  is given 
by~\footnote{More precise  numbers are available from the authors.}
\begin{eqnarray}
\left(
\begin{array}{llllllllll}
 \text{2.1} & -\text{1.58} & \text{0.02} & \text{0.} & \text{0.02} & \text{0.92} & \text{0.15} & \text{3.17} &
   \text{0.22} & \text{1.65} \\
 -\text{1.58} & \text{0.13} & \text{5.09} & \text{0.14} & -\text{0.33} & \text{3.51} & -\text{0.16} & -\text{0.35}
   & -\text{3.41} & \text{1.61} \\
 \text{0.02} & \text{5.09} & \text{0.83} & -\text{0.09} & \text{2.85} & -\text{0.97} & -\text{0.91} & \text{2.4} &
   -\text{0.25} & \text{4.45} \\
 \text{0.} & \text{0.14} & -\text{0.09} & -\text{4.19} & -\text{3.79} & \text{1.11} & -\text{1.03} & \text{0.45} &
   -\text{1.5} & \text{2.79} \\
 \text{0.02} & -\text{0.33} & \text{2.85} & -\text{3.79} & -\text{4.6} & -\text{0.12} & -\text{4.42} & \text{1.91}
   & -\text{0.45} & -\text{0.31} \\
 \text{0.92} & \text{3.51} & -\text{0.97} & \text{1.11} & -\text{0.12} & \text{0.25} & \text{0.35} & \text{0.41} &
   \text{0.74} & \text{2.58} \\
 \text{0.15} & -\text{0.16} & -\text{0.91} & -\text{1.03} & -\text{4.42} & \text{0.35} & -\text{3.07} &
   \text{2.61} & \text{2.24} & -\text{2.64} \\
 \text{3.17} & -\text{0.35} & \text{2.4} & \text{0.45} & \text{1.91} & \text{0.41} & \text{2.61} & \text{1.9} &
   \text{4.51} & -\text{0.15} \\
 \text{0.22} & -\text{3.41} & -\text{0.25} & -\text{1.5} & -\text{0.45} & \text{0.74} & \text{2.24} & \text{4.51}
   & \text{4.91} & \text{0.33} \\
 \text{1.65} & \text{1.61} & \text{4.45} & \text{2.79} & -\text{0.31} & \text{2.58} & -\text{2.64} & -\text{0.15}
   & \text{0.33} & \text{2.15}
\end{array}
\right)
&&\nonumber
\end{eqnarray}
in units of $-3.3 \times 10^{-16}$ GeV$^{-1}$, where we have assumed $\sin\beta \approx 1$. The corresponding Yukawa couplings $Y^T_\nu$  are 
\begin{eqnarray}
\left(
\begin{array}{llllllllll}
 \text{0.51} & \text{0.06} & -\text{0.32} & \text{0.32} & -\text{0.83} & -\text{0.36} & \text{1.23} & \text{0.41}
   & -\text{0.35} & \text{0.48} \\
 \text{0.37} & \text{0.1} & \text{1.36} & \text{0.62} & -\text{0.82} & -\text{0.66} & \text{0.06} & -\text{0.02} &
   -\text{0.78} & \text{0.11} \\
 -\text{0.09} & \text{0.89} & \text{0.13} & -\text{0.55} & \text{0.08} & -\text{0.25} & -\text{0.04} & \text{0.68}
   & \text{0.11} & -\text{1.1} 
\end{array}
\right).
&& \nonumber
\end{eqnarray}
The Majorana mass eigenvalues are $M_i=$(2.6, 2.7, 3.1, 4.0, 5.0, 8.1, 9.3, 10.5, 18, 58)$\times {10^{14}}$ GeV with the
geometric average scale 7.2$\times 10^{14}$ GeV. In the basis where the Majorana mass matrix is diagonal,
the Yukawa couplings remain of order one.

In the case of 2 right--handed neutrinos, we nevertheless use a 3$\times$3 matrix notation,
keeping in mind that $M^{-1}$ has rank two, so that only two right--handed neutrinos
contribute.  We take 
\begin{eqnarray}
&&M^{-1}=
\left(
\begin{array}{lll}
 \text{11.28} & -\text{8.82} & \text{21.86} \\
 -\text{8.82} & \text{6.91} & -\text{17.21} \\
 \text{21.86} & -\text{17.21} & \text{44.39}
\end{array}
\right)
\nonumber
\end{eqnarray}
in units of $-3.3 \times 10^{-16}$ GeV$^{-1}$, and 
\begin{eqnarray}
&& Y^T_\nu=
\left(
\begin{array}{lll}
 -\text{0.02} & -\text{1.19} & -\text{0.43} \\
 \text{0.83} & -\text{0.66} & -\text{0.88} \\
 \text{0.54} & \text{0.39} & \text{0.13}
\end{array}
\right)
.\nonumber
\end{eqnarray}
The Majorana mass eigenvalues are $M_i=$(0.5, 56)$\times 10^{14}$ GeV with the 
geometric average scale 5.2$\times 10^{14}$ GeV.

As expected,  the Majorana mass scale is somewhat higher in the first case, while the 
Yukawa couplings are similar in magnitude. For the comparison
of their lepton--flavour- and CP--violating
effects, the relevant quantity is $Y^{\dagger}_\nu L Y_\nu$,  where $L={\rm diag}[ \ln ( M_{\rm GUT} 
/M_i ) ]$. Ignoring the RG running of  $Y_\nu$ \cite{Antusch:2005gp}, we have 
\begin{eqnarray}
Y^{\dagger}_\nu L Y_\nu =
\left(
\begin{array}{lll}
 \text{7.47} & -\text{0.39} & -\text{1.65} \\
 -\text{0.39} & \text{9.77} & -\text{0.58} \\
 -\text{1.65} & -\text{0.58} & \text{8.4}
\end{array}
\right)
\nonumber
\end{eqnarray}
for the 10 right--handed neutrinos  case and 
\begin{eqnarray}
Y^{\dagger}_\nu L Y_\nu =
\left(
\begin{array}{lll}
 \text{0.67} & \text{1.19} & \text{0.15} \\
 \text{1.19} & \text{2.37} & \text{0.01} \\
 \text{0.15} & \text{0.01} & \text{0.29}
\end{array}
\right)
\nonumber
\end{eqnarray}
if there are  2 right--handed neutrinos. Here, as before, the (1,1) entry corresponds to the 
$\tau$-$\tau$ matrix element and so on. 

We find that not all of the entries are enhanced by the same factor in the 10 
right--handed neutrino case, but the phenomenological consequences are dramatic!
For example, the $\tau \rightarrow e\gamma$ branching ratio increases by a factor of
100 compared to the 2 right--handed neutrino case, while that for $\mu \rightarrow e\gamma$ is enhanced by a 
factor of $10^3$.
This result is of course specific to  the  above texture, however
it is generally true that the lepton--flavour--violating
transitions are more significant in the multi--neutrino case.
Although here we have considered a CP--conserving Ansatz,
similar statements apply in general to the EDMs as well. 
Since the magnitudes of the diagonal entries increase by factors of 10 or so,
the insertion  of CP phases of order unity would result in increases of EDMs by two orders
of magnitude.


The above example illustrates the main features of the multi--neutrino scenario. 
This particular, the 10 right--handed $\nu$ texture would be consistent with the 
BR($\mu \rightarrow e\gamma$) constraint only for slepton masses of $\sim 500$ GeV
or more. The corresponding predictions
for BR($\tau \rightarrow e\gamma$) and BR($\tau \rightarrow \mu \gamma$) are at the 
$10^{-10}-10^{-11}$
level. For low $\tan\beta$ and CP--violating phases of order unity, 
the electron EDM is expected to be of order $10^{-29}$ $e$--cm
and the muon EDM a factor of $10^2$ larger. Thus, the most promising observables for
this texture
would be  BR($\mu \rightarrow e\gamma$) and $d_e$ (see \cite{Semertzidis:2004uu} for 
a discussion of the experimental prospects).
These flavour--dependent effects would be  accentuated
further for more right--handed neutrinos.

\section{Comments on Leptogenesis}

A number of features of
leptogenesis  \cite{Fukugita:1986hr} 
with many right--handed neutrinos have been studied  by Eisele in  Ref.\cite{Eisele:2007ws}.
Here we discuss only a few of the most important differences of this scenario compared with 
the standard
scheme~\footnote{Recent reviews of leptogenesis can be found in Ref.\cite{Buchmuller:2005eh}.}.

The out--of--equilibrium decay of the lightest right--handed neutrino creates a lepton 
asymmetry
\begin{equation}
\eta_L = { n_{N}+n_{\tilde N} \over s} ~\epsilon ~ \delta \;,
\end{equation}
where $n$ and $s$ are the number and entropy densities, respectively; 
$\delta$ is the washout parameter  characterizing the fraction of the lepton
asymmetry surviving the washout effects, and $\epsilon$ is the 
CP--violating asymmetry:
\begin{equation}
\epsilon = { \Gamma - \overline \Gamma  \over \Gamma + \overline \Gamma  } \;,
\end{equation}
with $\Gamma$  being the decay rate of the singlet right--handed neutrino
into light leptons plus the Higgs boson,  and $\overline \Gamma$ being the decay rate
for the CP--conjugate process. In the case of hierarchical Majorana masses,
the CP asymmetry is dominated by the decays of the lightest singlet, and
\begin{equation}
\epsilon_1 \simeq -{3\over 8\pi}~ {M_1 \over  \langle \phi^u \rangle^2 } ~
{ {\rm Im}  \bigl[   Y_\nu M_{\rm eff}^\dagger Y_\nu^T        \bigr]_{11} \over
    \bigl[   Y_\nu  Y_\nu^\dagger        \bigr]_{11}   } \;,
\end{equation}
where $M_1$ is the lightest Majorana mass. 
Davidson and Ibarra  have derived an upper bound on $\epsilon_1$
in the $3\times 3$ case \cite{Davidson:2002qv}. 
Here we consider the corresponding bound in the multi--neutrino
scenario. Working in the basis where $M_{\rm eff}$ is diagonal and real,
we first note that the asymmetry is maximized for $Y_{1i}^2= {\rm i} \vert Y_{1i}\vert^2 $.
Then, the relevant  ratio satisfies
\begin{equation}
{ \sum_i m_i  \vert Y_{1i}\vert^2 \over  \sum_i   \vert Y_{1i}\vert^2 } \leq m_3 \;,
 \end{equation} 
where $m_3$ is the mass of the heaviest left--handed neutrino.
The maximum is achieved for $Y_{11}, Y_{12}=0$. Thus, we obtain a generalized bound 
for an arbitrary number of right--handed neutrinos:
\begin{equation}
\vert \epsilon_1 \vert \leq {3\over 8\pi}~ {M_1 \over  \langle \phi^u \rangle^2 } ~ m_3 \;.
\label{DI}
\end{equation}
This is identical to the ``weaker'' version of the Davidson--Ibarra bound in 
\cite{Davidson:2002qv}, which was also found in Ref.\cite{Asaka:1999yd}. 
For the $3\times 3$ case, a stronger bound is valid, which is
obtained from the above expression by replacing $m_3$ with $m_3-m_1$. This stronger
form does not apply to the $n\times n$ case, and one can construct examples 
with    $Y_{11,12}=0$,     Arg$(Y_{13})=\pi/4$ which saturate ({\ref{DI}}). 
Technically, this happens because the Casas--Ibarra parametrization  \cite{Casas:2001sr}   involving
$ orthogonal $  matrices does not apply to $n\times 3$ Yukawa 
couplings\footnote{The $n\times 3$  matrix $R$  entering the Casas--Ibarra parametrization
satisfies $R^TR={\bf 1}$ but not $R R^T={\bf 1}$.}. 

Thus, only the weaker version of the  Davidson--Ibarra bound holds in the general case.
Needless to say, this distinction is only relevant for a degenerate light--neutrino
spectrum. In the $3\times 3$ case, the bound on $\epsilon $ tightens as the overall scale
of light neutrino masses increases because $ m_3-m_1 \simeq \Delta m^2_{\rm atm}/ 2m_3$.
For many right--handed neutrinos, the only relevant bound is (\ref{DI}) which relaxes
as $m_3$ increases. This inequality can be interpreted as a lower bound on $M_1$, which
in turn implies a lower bound on the reheating temperature  $T_{\rm reh}$   \cite{Davidson:2002qv}.
If we take $m_3 \sim 0.5 $ eV, the   Davidson--Ibarra bound
on  $T_{\rm reh}$ relaxes by an order of magnitude\footnote{For systematic studies of bounds
on $T_{\rm reh}$, see \cite{Buchmuller:2002rq},\cite{wilfried}.    }, 
\begin{equation}  
T_{\rm reh} \geq 10^7-10^9~{\rm GeV} \;.
\end{equation}
This ameliorates somewhat  the problems of gravitino 
over--production~\cite{Khlopov:1984pf}  and moduli destabilization
at high temperature \cite{Buchmuller:2004xr}.

Another interesting consequence of the multi--$\nu_R$ scenario is the relaxation of the cosmological 
bound on  the lightest neutrino mass $m_1$. The requirement of out--of--equilibrium decay of the
lightest right--handed neutrino~\footnote{This requirement is lifted in the strong wash--out regime.
See \cite{Buchmuller:2003gz} for a discussion and the neutrino mass bounds. }  
 \cite{Fischler:1990gn},
$\Gamma_1 < H\vert_{T\simeq M_1}$, implies
\begin{equation}
\tilde m_1 = (Y_\nu Y_\nu^\dagger)_{11} { \langle  \phi^u  \rangle^2  \over M_1} \leq 5\times 10^{-3} ~ {\rm eV} \;.
 \end{equation}
In the $3\times 3$ case, $m_1 \leq \tilde m_1$ (see, for example, \cite{Fujii:2002jw},\cite{Davidson:2002qv}). 
In the multi--$\nu_R$  case, this is no longer true, since 
$m_1$ receives $n$ contributions, $m_1 \propto (Y_\nu^T Y_\nu /M_1)_{11}$, and one can easily have  
$m_1 > \tilde m_1$. Thus, the constraint on the lightest neutrino mass is relaxed.

We see that the presence of more than 3 right--handed neutrinos disrupts the usual relations
between the light neutrino masses and leptogenesis. In the above arguments, 
we have made the usual assumption that the lightest $\nu_R$ dominates in leptogenesis.
We have assumed  that the extra $\nu_R$'s play a passive role,
in the sense that they do not contribute to leptogenesis,
yet affect the light neutrino masses.

Further possible effects of many right--handed neutrinos have been discussed in  Ref.\cite{Eisele:2007ws}.
These include, for example, the possibility that more $\nu_R$'s are responsible for leptogenesis.
Such effects can be studied on a model-by-model basis.

\section{Conclusions}

We have explored some of the phenomenological features of the seesaw mechanism
with many right--handed neutrinos. Such a generalization is allowed by the absence of 
experimental  constraints on the number of 
heavy right--handed neutrinos, and is motivated by string 
constructions.
We find that, in such models, the Majorana mass scale can be significantly {\it higher} than that 
in the traditional scheme with 3 right--handed neutrinos, and have demonstrated 
with a specific example that
the magnitudes of  lepton--flavour-- and CP--violating transitions 
can be {\it enhanced} by orders of magnitude. 
We also find that certain constraints on leptogenesis are sensitive to the number 
of $\nu_R$ and can be relaxed. This applies, in particular, to 
 the bounds on the reheating temperature and the lightest
neutrino mass.

\section*{Acknowledgements.} 
O.L. is indebted to M.-T. Eisele, M. Ratz
and, especially, to A. Ibarra 
for enlightening discussions and useful correspondence.

\end{document}